\documentclass[12pt]{article}
\usepackage[english]{babel}
\usepackage{amsmath}
\usepackage{graphicx}
\usepackage{hyperref}
\usepackage{amsmath}
\usepackage{amsfonts}
\usepackage{amssymb}

\begin{document}

\title{\textbf{{}On Lagrangian Formulation for Spin $(2,1,1)$ Massive Tensor
Field on Minkowski Backgrounds}}
\author{\textsc{A.A. Reshetnyak${}^{a,b}$\thanks{%
reshet@ispms.tsc.ru \hspace{0.5cm} ${}^{\dagger }$moshin@rambler.ru}, P.Yu.
Moshin$^{c\dagger }$} \\
\textit{${}^{a}$ Laboratory of Computer-Aided Design of Materials,
Institute of }\\
\textit{Strength Physics and Materials Science, 634021 Tomsk, Russia}\\
[0.3cm] \textit{${}^b$Tomsk State Pedagogical University,}
\\\textit{634041 Tomsk, Russia }\\
[0.3cm] \textit{${}^c$Tomsk State University,} \textit{634051
Tomsk, Russia } }
\date{}
\maketitle

\begin{abstract}
We derive the Lagrangian for a new model of a massive rank-4
tensor field with generalized spin $(2,1,1)$ in Minkowski
spacetime of any dimension $d>5$, by using dimensional reduction
applied to a reducible gauge model of a massless rank-4 tensor
field with generalized spin $(2,1,1)$ in Minkowski spacetime of
dimension $d+1$.
\end{abstract}

\noindent \textsl{Keywords:} higher-spin fields; gauge theories;
Lagrangian formulation; higher-spin symmetry algebra.

\section{Introduction}

It is well known that massive and massless higher-spin fields are
high-energy excitations of superstring models \cite{HSstring},
including completely symmetric
\cite{Fronsdal,Fronsdal1,Vasiliev,massive
Minkowski,Vasiliev_mas,massive AdS} and mixed-symmetric
\cite{Labastida,Metsaev} fields of integer and half-integer spins.
These models imply the existence of interactions and particles
with spins higher than $2$, whose experimental discovery,
including the recent discovery of the Higgs boson, having the mass
of 125 GeV \cite{HiggsLHC}, indirectly confirming the  minimal
$N=1$ supesymmetric extension of the Standard Model (involving the
mass of a light Higgs boson), is one of the research tasks (for instance,
search for dark matter and supersymmetric partners) being carried
out at the LHC. Experimental search for these particles and
interactions calls for an adequate mathematical apparatus for
their description within the formalism of scattering amplitudes.

Lagrangian description for higher-spin fields is one of the basic
ingredients for computing the vacuum mean values of physical
quantities involved in high-energy processes within the Feynman
diagram technique using the Lagrangian quantum action provided by
the Batalin--Vilkovisky rules \cite{BV}. It should be noted that
the currently available literature (for a review, see, e.g.,
\cite{Vas_rev,Sorokin,Bouatta,Sagnotti,Bekaert,Tsulaia_rev})
contains quite a large number of results devoted to completely
symmetric and completely antisymmetric tensor and spin-tensor
fields in spaces of constant curvature in both the metric-like
\cite{Fronsdal,Fronsdal1} and the frame-like \cite{Vasiliev}
descriptions, whereas the case of mixed-symmetric higher-spin
fields has not been studied so well. Exceptions include
the recently constructed Lagrangian descriptions with traceless
constraints within the frame-like approach \cite{SkvortsovLF,framefermimix}
and the metric-like approach \cite{Franciamix} using the concept
of Bianchi identities in Minkowski spacetime, as well as
the (spin-)tensor fields subject to two-row \cite{Zinovievfermi,BRmixads}
and two-column \cite{Alkalaev} Young diagrams in anti-de Sitter
spaces.
Amongst the Lagrangian descriptions without constraints and with
specific values of generalized integer spin in the configuration
space involving no auxiliary fields, there have been known only
certain examples of fields described by two-row Young diagrams of
spin $\mathbf{s}=(2,1)$ in constant-curvature spaces
\cite{Vasilievmix}, $\mathbf{s}=(2,2)$ in Minkowski space
\cite{BurdikPashnev}, as well as the case of $4$-th rank massive tensors
in flat and (anti-)de Sitter spaces subject to the Young tableaux
$Y[3,1], Y[2,2]$ with two columns in terms of an antisymmetric basis
\cite{Zinoviev_mix_mass}, accompanied by a study of
the massless and partially massless limits.

The papers \cite{BRmixbos,mixboseResh,Rmixferm} of one of the
authors of the present work have been the first ones to obtain
unconstrained Lagrangian descriptions for arbitrary free particles
of integer and half-integer generalized spin on a basis of the
BRST (Becchi---Rouet---Stora--Tyutin) -- BFV
(Batalin--Fradkin--Vilkovisky) formalism, applied in its original
setting \cite{BFV} to the generalized canonical quantization of
arbitrary gauge theories with constraints, and reflecting the
presence of a special one-parameter supersymmetry, i.e., the
BRST-symmetry\footnote{In this respect, we recall that the first
results involving a constrained Lagrangian description for
completely symmetric bosonic fields have been presented in
\cite{Ouvry,Bengtsson}.} \cite{BRST}. In particular, the article
\cite{BRmixbos} was the first to obtain, by solving the
cohomological gauge complex, such a gauge-invariant Lagrangian
description for a massless $4$-th-rank tensor field $\Phi _{\mu
\nu ,\rho ,\sigma }$ of spin $(2,1,1)$ in a flat $d$-dimensional
spacetime $\mathbb{R}^{1,d-1}$ that is defined entirely in terms
of the tensor components of $\Phi _{\mu \nu ,\rho ,\sigma }$. At
the same time, no Lagrangian description for a free massive
$4$-th-rank tensor of spin $(2,1,1)$ in $\mathbb{R}^{1,d-1}$ has
been presented. Such a description is important due to the fact
that it may be able to provide a basis for constructing a
physically meaningful theory of interacting massive higher-spin
fields having no auxiliary fields and interacting with lower-spin
fields. In this respect, the present work
intends to find a Lagrangian for the model \cite{BRmixbos} of a massive $4$%
-th-rank tensor field of spin $(2,1,1)$. The construction may be carried out
in two different ways: either by repeating the procedure \cite{BRmixbos} of
solving the cohomological gauge complex in a special Fock space with the use
of a \emph{massive} tensor field $\Phi _{\mu \nu ,\rho ,\sigma }$, or by
applying dimensional reduction to an already known \emph{massless} gauge
model for a similar tensor field, determined, however, in a $(d+1)$-dimensional
flat spacetime. The present work intends to solve the above problem on a
basis of the latter approach.

The main part of the present work is organized as follows. First, we recall
the equations that express the fact that a massive field with generalized
spin $\mathbf{s}=(2,1,1)$ belongs to the relevant irreducible representation
of the Poincare group in terms of a mixed-symmetric $4$-th-rank tensor $\Phi
^{\mu \nu ,\rho ,\sigma }$ in a $d$-dimensional Minkowski spacetime. Second,
we present a gauge-invariant Lagrangian for a mixed-symmetric $4$-th-rank
tensor with generalized spin $\mathbf{s}=(2,1,1)$, defined in a $(d+1)$%
-dimensional Minkowski spacetime, and describe reducible gauge
transformations for this Lagrangian. Third, based on the rules of
dimensional reduction that transform a \emph{massless} theory in $\mathbb{R}%
^{1,d}$ into a \emph{massive} theory in $\mathbb{R}^{1,d-1}$,\ we apply these
rules to a Lagrangian model of a massless tensor field in $\mathbb{R}^{1,d}$.
We fix the reducible gauge invariance for a massive model, obtained by
eliminating the Stueckelberg auxiliary fields and lower-spin gauge
parameters, and make a transfer from a Lagrangian gauge model to a non-gauge
model, defined entirely in terms of a massive mixed-symmetric $4$-th-rank
tensor ${\Phi }_{\mu \nu ,[\rho ,\sigma ]}$. In Conclusion, we summarize the
obtained results and outline the perspectives of their application.

\section{Lagrangian Description of Massless Model in $\mathbb{R}^{1,d}$}

Let us recall that, based on the approach of E. Wigner, a massive particle
of generalized spin $\mathbf{s}=(2,1,1)$ is described within the metric
formalism by a $4$-th-rank tensor $\Phi ^{\mu \nu ,\rho ,\sigma }(x)\equiv
\Phi ^{\mu \nu ,\rho ,\sigma }$, being symmetric in its first two indices, $%
\Phi ^{\mu \nu ,\rho ,\sigma }=\Phi ^{\nu \mu ,\rho ,\sigma }$, and
satisfying the Klein--Gordon equation (\ref{Eq-0b}), as well as the
divergentless (\ref{Eq-1b}), traceless (\ref{Eq-2b}) and mixed-symmetry (\ref%
{Eq-3b}) equations:%
\begin{eqnarray}
&&(\partial ^{\mu }\partial _{\mu }+m^{2})\Phi _{\mu \nu ,\rho ,\sigma }=0,
\label{Eq-0b} \\
&&\partial ^{\mu }\Phi _{\mu \nu ,\rho ,\sigma }=\partial ^{\rho }\Phi _{\mu
\nu ,\rho ,\sigma }=\partial ^{\sigma }\Phi _{\mu \nu ,\rho ,\sigma }=0,
\label{Eq-1b} \\
&&\eta ^{\mu \nu }\Phi _{\mu \nu ,\rho ,\sigma }=\eta ^{\mu \rho }\Phi _{\mu
\nu ,\rho ,\sigma }=\eta ^{\mu \sigma }\Phi _{\mu \nu ,\rho ,\sigma }=\eta
^{\rho \sigma }\Phi _{\mu \nu ,\rho ,\sigma }=0,  \label{Eq-2b} \\
&&\Phi _{\{\mu \nu ,\rho \},\sigma }=0,\quad \Phi _{\{\mu \nu ,\hat{\rho}%
,\sigma \}}=0,\quad \Phi _{\mu \nu ,\{\rho ,\sigma \}}=0.  \label{Eq-3b}
\end{eqnarray}%
Eqs. (\ref{Eq-0b})--(\ref{Eq-3b}) reflect the property that the tensor field
$\Phi ^{\mu \nu ,\rho ,\sigma }$ belongs to an irreducible massive
representation of the Poincare group in $\mathbb{R}^{1,d-1}$, and, in
particular, Eqs. (\ref{Eq-3b}) imply that the tensor field $\Phi ^{\mu \nu
,\rho ,\sigma }$ belongs to the Young tableaux
$Y(2,1,1)=\begin{array}{|c|c|}
\hline
\!\mu \! & \!\nu \! \\ \cline{1-1}\hline
\!\rho \!  \\ \cline{1-1}
\!\sigma \!  \\  \cline{1-1}
\end{array}$,
whereas the last equation in (\ref{Eq-3b}) implies that it is only the
part of $\Phi _{\mu \nu ,\rho ,\sigma }$ which is antisymmetric, $\Phi _{\mu
\nu ,[\rho ,\sigma ]}=-\Phi _{\mu \nu ,[\sigma ,\rho ]}$, in the two final
indices $\rho ,\sigma $, that survives, where $\mu ,\nu ,\rho ,\sigma
=0,1,\ldots ,d-1$.

Our starting point is a gauge-invariant Lagrangian for a massless $4$%
-th-rank tensor $\Phi ^{MN,[R,S]}$ with generalized spin $\mathbf{s}=(2,1,1)$%
, defined in a $(d+1)$-dimensional Minkowski spacetime, being antisymmetric
with respect to the permutation of the third and fourth indices, $\Phi
^{MN,[R,S]}$= $-\Phi ^{MN,[S,R]}$, and subject to the (off-shell) Young
symmetry conditions%
\begin{equation}
\Phi ^{MN,[R,S]}+\Phi ^{RM,[N,S]}+\Phi ^{NR,[M,S]}=0,  \label{mixsymm}
\end{equation}%
which accompany equations of the form (\ref{Eq-0b})--(\ref{Eq-2b}), however,
in a $(d+1)$-dimensional spacetime, and also with $m=0$. For tensor
transformations, we use the mostly minus signature for the metric tensor $%
\eta _{MN}=\mathrm{diag}(+,-,...,-)$, with $(M,N,P,R,S)$ being $(d+1)$%
-dimensional Lorentz indices with the structure
\begin{equation}
(M,N,P,R,S)=\bigl((\mu ,d),(\nu ,d),(\pi ,d),(\rho ,d),(\sigma ,d)\bigr)%
,\quad \eta _{MN}=\left(
\begin{array}{cc}
\eta _{\mu \nu } & 0 \\
0 & -1%
\end{array}%
\right) ,  \label{MNPRS}
\end{equation}%
where $\mu ,\nu ,\pi ,\rho ,\sigma $ are $d$-dimensional Lorentz indices.

The action has the form \cite{BRmixbos}:
\begin{eqnarray}
\hspace{-1ex} &\hspace{-1ex}&\hspace{-1ex}\mathcal{S}_{(2,1,1)}\ =\ \int
d^{d+1}x\Bigl[\frac{1}{2}\Phi ^{MN,[R,S]}\Bigl\{\square \Phi
_{MN,[R,S]}+\partial _{\{M}\bigl[\textstyle\partial _{R}\Phi
^{T}{}_{[N\},[T,S]]}+{}\partial _{N\}}\Phi ^{T}{}_{T,[R,S]}  \label{S(2,1,1)}
\\
&\hspace{-1ex}&\hspace{-1ex}+\partial _{S}\Phi
^{T}{}_{[N\},[R],T]}-2\partial ^{T}\Phi _{N\}T,[R,S]}\bigr]+2\partial _{R}%
\bigl[\partial _{\{M}\Phi ^{T}{}_{T,[S,N\}]}+\partial _{\{N}\Phi
^{T}{}_{[M\},[S],T]}-2\partial ^{T}\Phi _{\{MT,[S,N\}]}\bigr]\Bigr\}  \notag
\\
&\hspace{-1ex}&\hspace{-1ex}-\frac{1}{4}\Phi _{T}{}^{T,[M,N]}\Bigl\{\square
\Phi ^{S}{}_{S,[M,N]}-2\partial _{\nu }\partial ^{R}\Bigl[\Phi
^{S}{}_{S,[R,M]}-\Phi ^{S}{}_{R,[S,M]}\Bigr]\Bigr\}  \notag \\
&\hspace{-1ex}&\hspace{-1ex}-2\Phi _{S}{}^{[M,[S,N]]}\partial ^{R}\partial
^{T}\Phi _{MT,[R,N]}-\Phi ^{S}{}_{S,}{}^{[M,N]}\partial ^{R}\partial
^{T}\Phi _{RT,[M,N]}+2\Phi ^{MN,[R,S]}\partial _{N}\partial ^{T}\Phi
_{MT,[R,S]}  \notag \\
&\hspace{-1ex}&\hspace{-1ex}+\frac{1}{2}\Phi
_{S}{}^{M}{}_{,}{}^{[S,N]}\partial _{M}\partial ^{R}\Bigl\{\Phi
^{T}{}_{R,[T,N]}-\Phi ^{T}{}_{T,[R,N]}\Bigr\}\Bigr].  \notag
\end{eqnarray}%
Due to relations (\ref{mixsymm}), the action (\ref{S(2,1,1)})
is invariant with respect to Abelian gauge transformations:%
\begin{equation}
\delta \Phi _{MN,[R,S]}=-\textstyle\frac{1}{2}\partial _{\{M}\xi
_{N\},[R,S]}+\frac{{1}}{2}\partial _{R}\xi _{\{M,[N\},S]}+\frac{{1}}{2}%
\partial _{S}\xi _{\{M,[\hat{R},N]\}}\;,  \label{gtr}
\end{equation}%
where the gauge parameters $\xi _{R,[M,N]}$ are defined
by the notation of \cite{BRmixbos} as $\phi _{1|R,[M,N]}^{(vi)1}$%
, and the symbol $\hat{}$ over the index $R$ in the function $\xi
_{\{M,[\hat{R},N]\}}$ implies that $R$ is not involved in
symmetrization, $\{M,[\hat{R},N]\}$,
defined by $\xi _{\{M,[\hat{R},N]\}}\equiv \xi _{M,[\hat{R},N]}+\xi _{N,[%
\hat{R},M]}$. In turn, the gauge transformations for the dependent gauge
parameters $\xi _{R,[M,N]}$, being antisymmetric under permutations of the
indices $M,N$ given by the rule $\xi _{R,[M,N]}\equiv \xi _{R,M,N}-\xi _{R,N,M}$,
and not belonging to any specific Young tableaux, have the form%
\begin{equation}
\delta \xi _{M,[N,R]}=2\partial _{\lbrack N}\xi _{M,R]}^{(1)}\mathtt{\ }%
\mathrm{and}\text{ }\delta \xi _{M,R}^{(1)}=-\partial _{R}\xi _{M}^{(2)},
\label{redgfinal}
\end{equation}%
where it has been taken into account that the first-stage gauge parameters
(second-rank tensor) $\xi _{M,R}^{(1)}$ are also gauge-dependent, which is
reflected in the presence of gauge transformations for them in terms of
(by now independent) second-stage gauge parameters $\xi _{M}^{(2)}$. By
means of the notation \cite{BRmixbos}, the quantities $\xi _{M,R}^{(1)},\xi
_{M}^{(2)}$ are defined as $\xi _{M,R}^{(1)}\equiv \phi _{1|M,R}^{\prime
\prime \prime 2}$ and $\xi _{M}^{(2)}\equiv \phi _{1|M}^{3}$. Relations (\ref%
{gtr}) and (\ref{redgfinal}) determine the model of a massless tensor field $%
\Phi ^{MN,[R,S]}$ with the action (\ref{S(2,1,1)}) as a gauge theory of a
massless tensor field of second-stage reducibility.

In view of (\ref{mixsymm}), the field tensor $\Phi _{MN,[R,S]}$
is doubly traceless. Notice that one may equivalently use the tensor
$\widehat{\Phi }_{MN,[R,S]}$,
\begin{equation}
\widehat{\Phi }_{MN,[R,S]}\equiv \Phi _{MN,[R,S]}-\frac{1}{2}\Phi _{\lbrack
RM,[N,S]]}-\frac{1}{2}\Phi _{N[R,[M,S]]},  \label{irrepYTensor}
\end{equation}%
which satisfies the Young symmetry conditions (\ref{mixsymm}) identically,
with the same action (\ref{S(2,1,1)}) and transformations (\ref{gtr}), after
a simple substitution $\Phi _{MN,[R,S]}\rightarrow \widehat{\Phi }%
_{MN,[R,S]} $.

\section{Lagrangian Description of Massive Model in $\mathbb{R}^{1,d-1}$}

First, the dimensional reduction of a massless theory in $\mathbb{R}^{1,d}$
to a massive one in $\mathbb{R}^{1,d-1}$ is given by the rules for
derivatives: $\partial _{M}=(\partial _{\mu },\imath m)$. Second, there
holds the following representation for the dynamical field and gauge
parameters:%
\begin{eqnarray}
\Phi _{MN,[R,S]} &=&\bigl({\Phi }_{\mu \nu ,[\rho ,\sigma ]},{\Phi }_{\mu
\nu ,[\rho ,d]},{\Phi }_{\mu {}d,[\rho ,\sigma ]},{\Phi }_{\mu {}d,[\rho
,d]},{\Phi }_{d{}d,[\rho ,\sigma ]},{\Phi }_{d{}d,[\rho ,d]}\bigr)\,,
\label{red211} \\
\xi _{M,[N,R]} &=&\bigl({\xi }_{\mu ,[\nu ,\rho ]},{\xi }_{\mu \lbrack \nu
,d]},{\xi }_{d,[\nu ,\rho ]},{\xi }_{d,[d,\rho ]}\bigr),  \label{red211g} \\
\xi _{M,N}^{(1)} &=&\bigl(\xi _{\mu ,\nu }^{(1)},\xi _{\mu ,d}^{(1)},\xi
_{d,\nu }^{(1)},\xi _{d,d}^{(1)}\bigr)\,,\qquad \xi _{M}^{(2)}\ =\ \bigl(\xi
_{\mu }^{(2)},\xi _{d}^{(2)}\bigr).  \label{red211gg}
\end{eqnarray}

The Young symmetry condition for the tensor $\Phi _{MN,[PS]}$ (\ref{mixsymm}%
) implies that after a projection onto $\mathbb{R}^{1,d-1}$ the scalar $\Phi
_{d{}d,[d,dS]}$ and the $d$-dimensional first-rank tensor ${\Phi }%
_{d{}d,[\rho ,d]}=\frac{1}{3}{\Phi }_{\{d{}d,[d\},\rho ]}$ vanish
identically, and there hold the following conditions for the remaining
tensor projections (\ref{red211}):\
\begin{equation}
{\Phi }_{d{}d,[\rho ,\sigma ]}\ =\ -2{\Phi }_{\rho {}d,[d,\sigma ]}\,,\quad {%
\Phi }_{\mu \nu ,[d,\rho ]}=-2{\Phi }_{d{}\{\mu ,[\nu \},\rho ]}\,,\quad {%
\Phi }_{\{\mu \nu ,[\rho \},\sigma ]}\ =\ 0.  \label{projPhi}
\end{equation}%
Therefore, the total configuration space of a massive free particle with
spin $\mathbf{s}=(2,1,1)$ in $\mathbb{R}^{1,d-1}$ contains one fourth-rank
massive mixed-symmetric tensor, ${\Phi }^{\mu \nu ,[\rho ,\sigma ]}$, an
auxiliary third-rank tensor, ${\varphi }_{\mu \nu ,\rho }\equiv {\Phi }_{\mu
\nu ,[d,\rho ]}$, and an auxiliary second-rank antisymmetric tensor, ${%
\varphi }_{[\rho \sigma ]}\equiv {\Phi }_{d{}d,[\rho ,\sigma ]}$. The two
latter tensors play the role of Stueckelberg fields.

The set of gauge parameters (\ref{red211g}) consists of one third-rank
tensor, ${\xi }_{\mu ,[\nu ,\rho ]}$, two antisymmetric second-rank tensors,
${\xi }_{[\nu ,\rho ]}\equiv {\xi }_{d,[\nu ,\rho ]}$ and ${\xi }_{\mu ,\nu
}\equiv {\xi }_{\mu \lbrack \nu ,d]}$, and one vector, $\xi _{\rho }\equiv {%
\xi }_{d,[d,\rho ]}$. In its turn, the set of first-stage gauge parameters (%
\ref{red211gg}) contains a second-rank tensor, $\xi _{\mu ,\nu }^{(1)}$, two
vectors, $(\xi _{\mu }^{(1)},\zeta _{\mu }^{(1)})\equiv (\xi _{\mu
,d}^{(1)},\xi _{d,\mu }^{(1)})$, and a scalar, $\xi ^{(1)}\equiv \xi
_{d,d}^{(1)}$. At the same time,\ the vector $\xi _{\mu }^{(2)}$ and scalar $%
\xi ^{(2)}\equiv \xi _{d}^{(2)}$ are second-stage gauge parameters.

The corresponding gauge-invariant action with the given auxiliary fields can
be deduced from (\ref{S(2,1,1)}) by dimensional reduction, $\mathbb{R}%
^{1,d}\rightarrow \mathbb{R}^{1,d-1}$, and must be invariant with respect to
the gauge transformations, implied by (\ref{gtr}),%
\begin{align}
& \delta {\Phi }_{\mu \nu ,[\rho ,\sigma ]}\hspace{1ex}=\hspace{1ex}-%
\textstyle\frac{1}{2}\partial _{\{\mu }\xi _{\nu \},[\rho ,\sigma ]}+\frac{{1%
}}{2}\partial _{\rho }\xi _{\{\mu ,[\nu \},\sigma ]}+\frac{{1}}{2}\partial
_{\sigma }\xi _{\{\mu ,[\hat{\rho},\nu ]\}}\,,  \label{phi211} \\
& \delta {\varphi }_{\mu \nu ,\rho }\hspace{1ex}=\hspace{1ex}\textstyle\frac{%
1}{2}\partial _{\{\mu }\xi _{\nu \},\rho }-\frac{{1}}{2}\partial _{\rho }\xi
_{\{\mu ,\nu \}}-\frac{{\imath }}{2}m\xi _{\{\mu ,[\hat{\rho},\nu ]\}}\,,
\label{varphi3} \\
& \delta {\varphi }_{[\rho \sigma ]}\hspace{1ex}=\hspace{1ex}-\textstyle{%
\imath }m\xi _{\lbrack \rho ,\sigma ]}\,,  \label{varphi2}
\end{align}%
which, in consequence of (\ref{redgfinal}),\ are reducible:
\begin{align}
& \delta \xi _{\mu ,[\nu ,\rho ]}=2\partial _{\lbrack \nu }\xi _{\mu ,\rho
]}^{(1)}, & & \delta \xi _{\lbrack \mu ,\nu ]}=2\partial _{\lbrack \mu
}\zeta _{\nu ]}^{(1)}\,, & &  \label{xi32} \\
& \delta \xi _{\mu ,\nu }\hspace{1ex}=\hspace{1ex}2\partial _{\nu }\xi _{\mu
}^{(1)}-2\imath m\xi _{\mu ,\nu }^{(1)}, & & \delta \xi _{\mu }\hspace{1ex}=%
\hspace{1ex}2\imath m\zeta _{\mu }^{(1)}-2\partial _{\mu }\xi ^{(1)}\,. & &
\label{xi1}
\end{align}%
At the same time, the second-stage reducibility implies%
\begin{align}
& \delta \xi _{\mu ,\nu }^{(1)}\hspace{1ex}=\hspace{1ex}-\partial _{\nu }\xi
_{\mu }^{(2)}\,, & & \delta \xi _{\mu }^{(1)}\hspace{1ex}=\hspace{1ex}%
-\imath m\xi _{\mu }^{(2)}\,,  \label{rxi1} \\
& \delta \zeta _{\mu }^{(1)}\hspace{1ex}=\hspace{1ex}-\partial _{\mu }\xi
^{(2)}\,, & & \delta \xi ^{(1)}\hspace{1ex}=\hspace{1ex}-\imath m\xi
^{(2)}\,.  \label{rxi12}
\end{align}

Transition to a non-gauge description is made by fixing the gauge
parameters $\xi _{\mu }^{(1)}$, $\xi ^{(1)}$ in (\ref{rxi1}), (\ref{rxi12})
with the help of shift transformations, with the corresponding independent
gauge parameters $\xi _{\mu }^{(2)}$, $\xi ^{(2)}$ of second stage, so that
there arises an intermediate reducible gauge theory of first-stage
reducibility with independent gauge parameters $\xi _{\mu ,\nu }^{(1)}$, $%
\zeta _{\mu }^{(1)}$. In a similar way, one can remove the gauge tensors $%
\xi _{\mu ,\nu }$, $\xi _{\mu }$ by using respective gauge shifts with
the help of $\xi _{\mu ,\nu }^{(1)}$, $\zeta _{\mu }^{(1)}$ in (\ref{xi1}).
In the gauge theory with the remaining independent gauge parameters
$\xi _{\mu ,[\nu ,\rho ]}$, $\xi _{\lbrack \mu ,\nu ]}$ one finally
eliminates the fields ${\varphi }_{\mu \nu ,\rho }$, ${\varphi }_{[\rho \sigma ]}$
in relations (\ref{varphi3}), (\ref{varphi2}) by using gauge shift
transformations with the respective parameters $\xi _{\mu ,[\nu ,\rho ]}$, $\xi
_{\lbrack \mu ,\nu ]}$, so that the theory becomes a non-gauge
one, defined entirely in terms of the massive mixed-symmetric $4$-th-rank
tensor ${\Phi }_{\mu \nu ,[\rho ,\sigma ]}$.

As a result, the Lagrangian of a massive $4$-th-rank tensor $\Phi ^{\mu \nu
,[\rho ,\sigma ]}$ with generalized spin $\mathbf{s}=(2,1,1)$ in a $d$%
-dimensional Minkowski spacetime, for the tensor $\widehat{\Phi }^{\mu \nu
,[\rho ,\sigma ]}$\ subject to conditions (\ref{irrepYTensor}), acquires the
form
\begin{eqnarray}
\mathcal{L}_{(2,1,1)} &=&\frac{1}{2}\widehat{\Phi }^{\mu \nu ,[\rho ,\sigma
]}\Bigl\{(\square +m^{2})\widehat{\Phi }_{\mu \nu ,[\rho ,\sigma ]}+\partial
_{\{\mu }\bigl[\textstyle\partial _{\rho }\widehat{\Phi }^{\tau }{}_{[\nu
\},[\tau ,\sigma ]]}+{}\partial _{\nu \}}\widehat{\Phi }^{\tau }{}_{\tau
,[\rho ,\sigma ]}+\partial _{\sigma }\widehat{\Phi }^{\tau }{}_{[\nu
\},[\rho ],\tau ]}\notag  \\
&&-2\partial ^{\tau }\widehat{\Phi }_{\nu \}\tau ,[\rho ,\sigma ]}\bigr]%
+2\partial _{\rho }\bigl[\partial _{\{\mu }\widehat{\Phi }^{\tau }{}_{\tau
,[\sigma ,\nu \}]}+\partial _{\{\nu }\widehat{\Phi }^{\tau }{}_{[\mu
\},[\sigma ],\tau ]}-2\partial ^{\tau }\widehat{\Phi }_{\{\mu \tau ,[\sigma
,\nu \}]}\bigr]\Bigr\}  \notag \\
&&-\frac{1}{4}\widehat{\Phi }_{\tau }{}^{\tau ,[\mu ,\nu ]}\Bigl\{(\square
+m^{2})\widehat{\Phi }^{\sigma }{}_{\sigma ,[\mu ,\nu ]}-2\partial _{\nu
}\partial ^{\rho }\Bigl[\widehat{\Phi }^{\sigma }{}_{\sigma ,[\rho ,\mu ]}-%
\widehat{\Phi }^{\sigma }{}_{\rho ,[\sigma ,\mu ]}\Bigr]\Bigr\}  \notag \\
&&-2\widehat{\Phi }_{\sigma }{}^{[\mu ,[\sigma ,\nu ]]}\partial ^{\rho
}\partial ^{\tau }\widehat{\Phi }_{\mu \tau ,[\rho ,\nu ]}-\widehat{\Phi }%
^{\sigma }{}_{\sigma ,}{}^{[\mu ,\nu ]}\partial ^{\rho }\partial ^{\tau }%
\widehat{\Phi }_{\rho \tau ,[\mu ,\nu ]}+2\widehat{\Phi }^{\mu \nu ,[\rho
,\sigma ]}\partial _{\nu }\partial ^{\tau }\widehat{\Phi }_{\mu \tau ,[\rho
,\sigma ]}  \notag \\
&&+\frac{1}{2}\widehat{\Phi }_{\sigma }{}^{\mu }{}_{,}{}^{[\sigma ,\nu
]}\partial _{\mu }\partial ^{\rho }\Bigl\{\widehat{\Phi }^{\tau }{}_{\rho
,[\tau ,\nu ]}-\widehat{\Phi }^{\tau }{}_{\tau ,[\rho ,\nu ]}\Bigr\}.  \label{S(2,1,1)mass}
\end{eqnarray}%
%
%
%
%
%
%
%
%
%
%
%
%
%


\section{Conclusion}

In this work, we have obtained a new Lagrangian for a free particle of mass $%
m$ and spin $\mathbf{s}=(2,1,1)$ in a $d$-dimensional Minkowski spacetime,
by applying dimensional reduction to a gauge-invariant model for a massless
free particle of the same spin in a $(d+1)$-dimensional Minkowski spacetime,
which has been obtained earlier on a basis of the universal BRST--BFV
approach in \cite{BRmixbos}. It should be noted that the resulting
Lagrangian is applicable to the dimensions $d\geq 6$ \cite{BRmixbos,mixboseResh}
subject to the inequality $d\leq 10$, implied by string theory.

In future perspective, we find it promising to use the respective Lagrangian
descriptions for massive and massless free particles as initial models
for the construction of interacting Lagrangian theories, in particular,\
those interacting with an external electromagnetic field. We also suggest
extending the Lagrangian description to the case of spin $\mathbf{s}%
=(2,1,\ldots ,1)$ with $(k-1)$ projections of generalized spin equal to $1$,
on condition that $k\leq \lbrack {d}/{2}]$ \cite{BRmixbos}.

\section*{Acknowledgments}

The authors are grateful to I.L. Buchbinder for stimulating
discussions.  A.A.R.  thanks Yu.M. Zinoviev for helpful remarks,
in particular, for pointing out the results of \cite{Zinoviev_mix_mass}.

A.A.R. thanks  the Galileo Galilei Institute for Theoretical Physics for the hospitality and the INFN for partial support during the completion of this work. The study was supported by the Ministry of Education and Science
of the Russian Federation, projects 14.B37.21.0774 and
14.B37.21.1301. The work of A.A.R. was partially supported by the
RFBR grant, project No. 12-02-000121, and by the grant of Leading
Scientific Schools of the Russian Federation, project No.
224.2012.2.


\begin{thebibliography}{99}
\bibitem{HSstring} A.~Sagnotti, M.~Tsulaia, \textit{On higher spins and the
tensionless limit of String Theory}, Nucl. Phys. B682 (2004) 83--116,
[arXiv:hep-th/0311257].

\bibitem{Fronsdal} C.~Fronsdal, \textit{Massless fields with integer spin},
Phys. Rev. D18 (1978) 3624--3629.

\bibitem{Fronsdal1} J.~Fang, C.~Fronsdal, \textit{Massless fileds with
half-integral spin}, Phys. Rev. D18 (1978) 3630--3633.

\bibitem{Vasiliev} M.A. Vasiliev,\textit{\ 'Gauge' form of description of
massless fields with arbitrary spin}, Sov. J. Nucl. Phys. 32 (1980) 439--445;
Yad. Fiz. 32 (1980) 855--861.

\bibitem{massive Minkowski}L.P.S.~Singh, C.R.~Hagen, \textit{Lagrangian formulation for arbitrary
spin. 1. The bosonic case}, Phys. Rev. D9 (1974) 898--909;
\textit{Lagrangian formulation for arbitrary spin. 2. The fermionic case},
Phys. Rev. D9 (1974) 910--920.

\bibitem{Vasiliev_mas}D.S.~Ponomarev, M.A.~Vasiliev,
\textit{Frame-Like Action and Unfolded Formulation for Massive Higher-Spin
Fields},  Nucl. Phys. B839 (2010) 466--498,
[arXiv:1001.0062[hep-th]].

\bibitem{massive AdS}Yu.M.~Zinoviev, \textit{On massive high spin particles in
AdS}, [arXiv:hep-th/0108192]; R.R.~Metsaev, \textit{Massive
totally symmetric fields in AdS(d)}, Phys.Lett. B590 (2004)
95--104, [arXiv:hep-th/0312297]; \textit{Fermionic fields in the
d-dimensional anti-de Sitter spacetime}, Phys. Lett. B419 (1998)
49--56, [arXiv:hep-th/9802097]; \textit{Light-cone form of field
dynamics in anti-de Sitter space-time and AdS/CFT correspondence},
Nucl. Phys. B563 (1999) 295--348, [arXiv:hep-th/9906217];
\textit{Massless arbitrary spin fields in AdS(5)}, Phys. Lett.
B531 (2002) 152--160, [arXiv:hep-th/0201226];  P.~de~Medeiros,
\textit{Massive gauge-invariant field theories on space of
constant curvature}, Class. Quant. Grav. 21 (2004) 2571--2593,
[arXiv:hep-th/0311254].

\bibitem{Labastida} J.M.F. Labastida,\textit{\ Massless fermionic free fields%
}, Phys. Lett. B186 (1987) 365--369; \textit{Massless bosonic free fields},
Phys. Rev. Lett. 58 (1987) 531--534; \textit{Massless particles in arbitrary
representations of the Lorentz group}, Nucl. Phys. B322 (1989) 185--209.

\bibitem{Metsaev} R.R.~Metsaev,\textit{\ Massless mixed symmetry bosonic
free fields in d-dimensional anti-de Sitter space-time}, Phys. Lett. B354
(1995) 78--84.

\bibitem{HiggsLHC} ATLAS Collaboration et al, \textit{Observation of a new
particle in the search for the Standard Model Higgs boson with the
ATLAS detector at the LHC}, Phys. Lett. B716 (2012) 1--29; CMS
collaboration et al, \textit{Observation of a new boson at a mass
of 125 GeV with the CMS experiment at the LHC}, Phys. Lett. B716
(2012) 30--61.

\bibitem{BV} I.A. Batalin, G.A. Vilkovisky, \textit{Gauge algebra and
quantization}, Phys. Lett. B102 (1981) 27--31; I.A. Batalin, G.A.
Vilkovisky, \textit{Quantization of gauge theories with linearly
dependent generators}, Phys. Rev. D28 (1983) 2567--2582.


\bibitem{Vas_rev} M.~Vasiliev, \textit{Higher spin gauge theories in various
dimensions}, Fortsch. Phys. 52 (2004) 702--717, [arXiv:hep-th/0401177].

\bibitem{Sorokin} D.~Sorokin, \textit{Introduction to the classical theory of higher spins},
AIP Conf. Proc. 767 (2005) 172--202, [arXiv:hep-th/0405069].

\bibitem{Bouatta} N.~Bouatta, G.~Comp\`ere,  A.~Sagnotti, \textit{An introduction to free
higher-spin fields}, [arXiv:hep-th/0409068].

\bibitem{Sagnotti} A.~Sagnotti,
E.~Sezgin, P.~Sundell, \textit{On higher spins with a strong Sp(2,R)
sondition}, [arXiv:hep-th/0501156].

\bibitem{Bekaert} X.~Bekaert, S.~Cnockaert,
C.~Iazeolla, M.A.~Vasiliev, \textit{Nonlinear higher spin theories in
various dimensions}, [arXiv:hep-th/0503128].

\bibitem{Tsulaia_rev} A.~Fotopoulos,
M.~Tsulaia, \textit{Gauge Invariant Lagrangians for Free and
Interacting Higher Spin Fields. A review of BRST formulation},
Int. J. Mod. Phys. A24 (2008) 1--60, [arXiv:0805.1346[hep-th]].

\bibitem{SkvortsovLF}E.D.~Skvortsov,
\textit{Frame-like Actions for Massless Mixed-Symmetry Fields in Minkowski
space}, Nucl. Phys. B808 (2009) 569, [arXiv:0807.0903[hep-th]].

\bibitem{framefermimix} E.D.~Skvortsov, Yu.M.~Zinoviev, \textit{Frame-like
Actions for Massless Mixed-Symmetry Fields in Minkowski space.
Fermions}, Nucl. Phys. B843 (2011) 559--569,
[arXiv:1007.4944[hep-th]].

\bibitem{Franciamix}
A.~Campoleoni, D.~Francia, J.~Mourad, A.~Sagnotti,
\textit{Unconstrained Higher Spins of Mixed Symmetry. I. Bose
Fields}, Nucl.Phys. B815 (2009) 289--357,
[arXiv:0810.4350[hep-th]].

\bibitem{Zinovievfermi} Yu.M.~Zinoviev, \textit{Toward frame-like gauge invariant formulation
for massive mixed symmetry bosonic fields}, Nucl. Phys. B812
(2009) 46--63, [arXiv:0809.3287[hep-th]].

\bibitem{BRmixads} C. Burdik, A. Reshetnyak, \textit{On representations of
Higher Spin symmetry algebras for mixed-symmetry HS fields on AdS-spaces.
Lagrangian formulation}, J. Phys. Conf. Ser. 343 (2012) 012102,
[arXiv:1111.5516[hep-th]].

\bibitem{Alkalaev} K.B. Alkalaev,\textit{Two-column higher spin massless
fields in AdS(d)}, Theor. Math. Phys. 140 (2004) 1253--1263; Teor. Mat. Fiz.
140 (2004) 424--436, [arXiv:hep-th/0311212].

\bibitem{Vasilievmix} L.~Brink, R.R.~Metsaev, M.A.~Vasiliev, \textit{How
massless are massless fields in $AdS_d$}, Nucl. Phys. B586 (2000) 183--205,
[arXiv:hep-th/0005136].

\bibitem{BurdikPashnev} C.~Burdik, A.~Pashnev, M.~Tsulaia, \textit{On the
mixed symmetry irreducible representations of the Poincare group in the BRST
approach}, Mod. Phys. Lett. A16 (2001) 731--746, [arXiv:hep-th/0101201].

\bibitem{Zinoviev_mix_mass} Yu.M. Zinoviev,
\textit{On Massive Mixed Symmetry Tensor Fields in Minkowski Space and (A)dS},
[arXiv:hep-th/0211233].

\bibitem{BRmixbos} I.L. Buchbinder, A. Reshetnyak, \textit{General
Lagrangian Formulation for Higher Spin Fields with Arbitrary Index Symmetry.
I. Bosonic fields}, Nucl. Phys. B862 (2012) 270--327,
[arXiv:1110.5044[hep-th]].

\bibitem{mixboseResh} A.A. Reshetnyak, \textit{On Lagrangian formulations
for arbitrary bosonic HS fields on Minkowski backgrounds}, Phys.
of Particles and Nuclei 43 (2012) 689--693, [arXiv:1202.3859[hep-th]].

\bibitem{Rmixferm} A.A. Reshetnyak, \textit{General Lagrangian Formulation
for Higher Spin Fields with Arbitrary Index Symmetry. 2. Fermionic fields},
Nucl. Phys. B869 (2013) 523--597, [arXiv:1211.1273[hep-th]].

\bibitem{BFV} E.S. Fradkin, G.A. Vilkovisky, \textit{Quantization of
relativistic systems with constraints}, Phys. Lett. B55 (1975)
224--226; I.A. Batalin, G.A. Vilkovisky, \textit{Relativistic
S-matrix of dynamical systems with boson and fermion constraints},
Phys. Lett. B69 (1977) 309--312; I.A. Batalin, E.S. Fradkin,
\textit{Operator quantization of relativistic dynamical systems
subject to first class constraints}, Phys. Lett. B128 (1983)
303--308.

\bibitem{BRST} I.V. Tyutin, \textit{Gauge invariance in field theory and
statistical physics in operator formalism}, Lebedev preprint FIAN
No. 39 (1975), [arXiv:0812.0580[hep-th]]; C. Becchi, A. Rouet, R. Stora, \textit{%
Renormalization of the abelian Higgs-Kibble model}, Commun. Math.
Phys. 42 (1975) 127--162.

\bibitem{Ouvry}S. Ouvry, J. Stern, \textit{Gauge fields of any spin and
symmetry}, Phys. Lett. B177 (1986) 335--340.

\bibitem{Bengtsson}A.K.H. Bengtsson, \textit{A
unified action for higher spin gauge bosons from covariant string
theory}, Phys. Lett. B182 (1986) 321--325.
\end{thebibliography}
\end{document}